\documentclass[a4paper,12pt]{article}
\usepackage[a4paper, total={6in, 8in}]{geometry}
\usepackage[utf8]{inputenc}
\usepackage{amsmath}
\usepackage{amsthm}
\usepackage{bbm}
\usepackage{amsfonts}
\usepackage{amssymb}
\usepackage{enumerate}
\usepackage{hyperref}
\usepackage{natbib}

\title{An extension to the applicability of adaptive bandwidth choice}
\date{\today}
\author{Dehao Dai\footnote{Department of Mathematics, Univ.~of California, San Diego;  email: ddai@ucsd.edu}}
\begin{document}

\maketitle
\begin{abstract}
    
    In this paper, we extend the applicability of the bandwidth choice method of \citet{Politis2003} by relaxing the conditions of his Theorem 2.3.
\end{abstract}

\section{Introduction}
Assume $X_{1}, \cdots, X_{N}$ are data from the strictly stationary time series $\{X_n, n\in \mathbb{Z}\}$ with mean $\mu=\mathbb{E} X_{t}$, and autocovariance $\gamma(k)=\mathbb{E}\left(X_{t}-\mu\right)\left(X_{t+|k|}-\right.$ $\mu)$; here both $\mu$ and $\gamma(\cdot)$ are unknown. Now we consider the problem of estimating the spectral density function $f(\omega) = (2 \pi)^{-1} \sum_{k=-\infty}^{\infty} e^{i k\omega } \gamma(k)$, for $\omega \in [-\pi, \pi]$ with kernel $\Lambda_h$. The corresponding kernel estimator may be written as:
$$
\hat{f}(\omega)=\Lambda_h * I_{N}(\omega)
$$
where $I_{N}(\omega)=(2 \pi)^{-1} \sum_{k=-N+1}^{N-1} e^{i k \omega} \hat{\gamma}(k)$ is the periodogram and $*$ denotes convolution; here, the real number $h$ is the bandwidth. In \citet{Politis2003}, the kernel $\Lambda_h(\omega)$ is defined as
\begin{equation*}
\Lambda_h(\omega) =\frac{1}{2\pi} \sum_{k =-M}^M \lambda^T(k/M)e^{ik\omega}
\end{equation*}
where $h = 1/M$ is regarded as the bandwidth and the function $\lambda^T(\cdot)$ has a trapezoidal shape symmetric around zero, i.e.
$$
\lambda^T(t) = \begin{cases}
 1 & \text{ if } |t| \in [0,c] \\
 \frac{1}{1-c}(1-|t|)  & \text{ if } t\in[c,1] \\
 0 & \text{ else}.
\end{cases}
$$
Note that $\lambda^T$ depends on the breaking point $c$.

An alternative form to represent $\hat{f}(w)$ in \citet{Politis2003} is the Fourier transform of the kernel, namely $\lambda_h(k) = \lambda^T(hk)$
$$
\hat{f}(w)=(2 \pi)^{-1} \sum_{k=-\infty}^{\infty} e^{i w k} \lambda^T(kh) \hat{\gamma}(k)
$$
where $\hat{\gamma}(k)=N^{-1} \sum_{i=1}^{N-|k|}\left(X_{i}-\bar{X}_{N}\right)\left(X_{i+|k|}-\bar{X}_{N}\right)$ is the lag-$k$ sample autocovariance for $|k|<N$ and  $\hat{\gamma}(k)$ is defined to be zero for $|k| \geq N$.

\citet{Politis2003} proposed the following  empirical rule of picking $M$ with $c = 1/2$.

\textbf{EMPIRICAL RULE OF PICKING $M$:}   \textit{ Let $\rho(k)= \gamma(k) / \gamma(0)$, and $\hat{\rho}(k)=\hat{\gamma}(k) / \hat{\gamma}(0)$. Let $\hat{m}$ be the smallest positive integer such that $|\hat{\rho}(\hat{m}+k)|<c \sqrt{\log N /N}$, for $k=1, \ldots, K_{N}$, where $c>0$ is a fixed constant, and $K_{N}$ is a positive, nondecreasing integer-valued function of $N$ such that $K_{N}=o(\log N)$. Then, let $\hat{M}=\hat{m}/c = 2 \hat{m}$.} 

Theorem 2.2 of \citet{Politis2003} showed that the above empirical rule
performs well in three cases for $\rho(k)$, namely (i) polynomial decay, 
(ii) exponential decay, and (iii) hard cut-off. 
In the next section, we complement his Theorem 2.2 by providing two weaker formulations of cases (i) and (ii).  

\section{Main result}

\textbf{Corollary}: 
Assume conditions strong enough to ensure that for all finite $n$,
\begin{equation}\label{eq-1}
\max _{k=1, \ldots, n}|\hat{\rho}(s+k)-\rho(s+k)|=O_\mathbb{P}(1 / \sqrt{N})
\end{equation}
uniformly in $s$, and
\begin{equation}\label{eq-2}
\max_{k=0,1, \ldots, n-1}|\hat{\rho}(k)-\rho(k)|=O_\mathbb{P}\left(\sqrt{\frac{\log N}{N}}\right).
\end{equation}
Also assume that the sequence $\gamma(k)$ does not have more than $K_N-1$ zeros in its first $k_{0}$ lags (i.e., for $\left.k=1, \ldots, k_{0}\right)$.
\begin{enumerate}[(i)]
\item Assume that $\gamma(k)= \sum_{i = 1}^l C_i k^{-d_i} \cos (a_i k+\theta_i)$ for $k>k_{0}$, and for some $C_i>0$ and positive integers $d_i$ and $l$, and some distinct constants $a_i \geq \frac{\pi}{K_N}, \theta_i \in[0,2 \pi]$ for all $i = 1, \cdots, l$. Then,
$$
\hat{M} \stackrel{P}{\sim} A_1 \left(\frac{N}{\log N} \right)^{1 / 2 d}
$$
for some positive constant $A_{1}$, and $d = \min_{i \in [l]}\{d_i \} $.
\item Assume $\gamma(k)= \sum_{i = 1}^l C_i \xi_i^{k} \cos (a_i k+\theta_i)$ for $k>k_{0}$, where $C>0,|\xi|<1, a_i \geq \frac{\pi}{K_N}, \theta_i \in[0,2 \pi]$ and positive integer $l$ are some constants. Then,
$$
\hat{M} \stackrel{P}{\sim} A_{2} \log N
$$
where $A_{2}=-1 / \log |\xi|$ with $|\xi| = \max_{i \in [l]} \{|\xi_i|\} $.

\end{enumerate}

\textbf{Proof} of (i): First note that from the condition that the sequence $\gamma(k)$ does not have more that $K_N - 1$ zeros in its first $k_0$ lags with (\ref{eq-1}), it follows that $\hat{m}>k_0$ with high probability, so we can focus on the part of the correlogram to the right of $k_0$, i.e., look at $\hat{\gamma}(k)$ for $k>k_0$ only.

Now condition $a_i \geq \pi/K_N$ implies that $K_N \geq \max\{ \pi /a_i\} $, i.e. $K_N$ is bigger than all the half-periods of the $\cos (a_i k +\theta)$ for $i = 1,\cdots, l$. Hence, (\ref{eq-2}) implies that
\begin{equation} \label{eq-3}
\max_{k = 1,\cdots,K_N} |\hat{\rho}(\hat{m}+k)| \geq C {(\hat{m}+K_N)}^{-d} +O_\mathbb{P}(\sqrt{\log N/N}),
\end{equation}
where $d = \min_{i \in [l]} \{d_i \}$, since
$$
\sum_{i = 1}^l C_i k^{-d_i} =Ck^{-d}  \left(\frac{C_L}{C}+\sum_{i = 1, i\neq L}^l\frac{C_i}{C}k^{d-d_i}\right) \geq Ck^{-d}
$$
where $L = \arg \min_{i \in [l]} \{d_i\}$ and $C = \max_{i \in [l]} \{C_i \}$.

From part (i) assumption with (\ref{eq-2}), it follows that 
$$
|\hat{\rho}(k)| = |Ck^{-d}| + O_\mathbb{P}(\sqrt{\log N/N}) \quad \text{uniformly in k},
$$
where $d = \min_{i \in [l]} \{d_i\}$, 
i.e.,
\begin{equation}\label{eq-4}
    |\hat{\rho}(\hat{m})| = |C\hat{m}^{-d}| + O_\mathbb{P}(\sqrt{\log N/N}).
\end{equation}

The empirical rule for picking $M$ implies that $|\hat{\rho}(\hat{m})| \geq c \sqrt{\log N/N}$, whereas
$$
\max_{k = 1, \cdots, K_N} |\hat{\rho}(\hat{m}+k)|<c\sqrt{\log N/N}.
$$

Thus, the above two statements with (\ref{eq-3}) and (\ref{eq-4})
$$
A_1 \left(\frac{N}{\log N}\right)^{1/2d} - K_N \leq \hat{m} \leq A_1 \left(\frac{N}{\log N}\right)^{1/2d} 
$$
with high probability. Since $K_N = o(\log N)$, 
$$
\hat{M} \stackrel{P}{\sim} A_1 \left(\frac{N}{\log N} \right)^{1 / 2 d},
$$
where $A_1$ is some positive constant and $d = \min_{i \in [l]} \{d_i \}$.

\textbf{Proof} of (ii): It is similar to proof of part (i). First, Hence, (\ref{eq-2}) implies that
\begin{equation} \label{eq-5}
\max_{k = 1,\cdots,K_N} |\hat{\rho}(\hat{m}+k)| \geq C |\xi|^{\hat{m}+K_N} +O_\mathbb{P}(\sqrt{\log N/N}),
\end{equation}
where $|\xi| = \max_{i \in [l]} \{|\xi_i| \}$, since
$$
\sum_{i = 1}^l C_i|\xi_i|^k = C|\xi|^k \left(\frac{C_L}{C} + \sum_{i = 1, i\neq L}^l \frac{C_i}{C}\left|\frac{\xi_i}{\xi}\right|^k\right) \geq C|\xi|^k
$$
where $L = \arg \max_i \{|\xi_i|\}$ and $C = \max_{i \in [l]} \{C_i \}$.

From part (ii) assumption with (\ref{eq-2}), it follows that 
$$
|\hat{\rho}(k)| = |C\xi^k| + O_\mathbb{P}(\sqrt{\log N/N}) \quad \text{uniformly in k},
$$
where $|\xi| = \max_{i \in [l]} \{|\xi_i| \}$,
i.e.,
\begin{equation}\label{eq-6}
    |\hat{\rho}(\hat{m})| = |C\xi^{\hat{m}}| + O_\mathbb{P}(\sqrt{\log N/N}).
\end{equation}

Thus, the two statements of the empirical rule for picking $M$ mentioned in part (i) with (\ref{eq-5}) and (\ref{eq-6}) implies
$$
A_2 \log N  - K_N \leq \hat{m} \leq A_2 \log N 
$$
with high probability. Thus,
$$
\hat{M} \stackrel{P}{\sim} A_2 \log N
$$
where $A_{2}=-1 / \log |\xi|$ with $|\xi| = \max_{i \in [l]} \{|\xi_i|\} $.

\vskip .1in
\noindent
Note that a general sufficient condition guaranteeing eq. (\ref{eq-1}) and (\ref{eq-2})
has been given by \citet{xiao2012covariance}.
\bibliography{main}
\nocite{politis_2011}
\end{document}